\documentclass{PoS}

\newcommand{\simge}{\hspace*{0.2em}\raisebox{0.5ex}{$>$}
     \hspace{-0.8em}\raisebox{-0.3em}{$\sim$}\hspace*{0.2em}}
\newcommand{\simle}{\hspace*{0.2em}\raisebox{0.5ex}{$<$}
     \hspace{-0.8em}\raisebox{-0.3em}{$\sim$}\hspace*{0.2em}}

\title{Nuclear Physics from QCD}

\ShortTitle{Nuclear Physics from QCD}

\author{\speaker{U. van Kolck}
\thanks{Research supported by the U.S. Department of Energy.}
\\
        Department of Physics, University of Arizona, Tucson, AZ 85745, USA\\
        E-mail: \email{vankolck@physics.arizona.edu}}

\abstract{
Effective field theories provide a bridge between QCD and nuclear physics.
I discuss light nuclei from this perspective,
emphasizing the role of fine-tuning.
}

\FullConference{8th Conference Quark Confinement and the Hadron Spectrum \\
		 September 1-6 2008\\
		 Mainz, Germany}

\begin{document}

\section{Introduction}
Nuclear physics has long been recognized as rich and complex.
The origin of this complexity has, however, been more difficult to pinpoint.
About fifteen years ago a road-map was traced \cite{origins}
to link nuclear physics to QCD using effective field theory (EFT).
The idea is to break the problem into two steps:
one is to calculate nuclear observables from a low-energy EFT;
another is to obtain the EFT from QCD.
The second step, where great progress has been achieved recently
through lattice simulations, 
was addressed  \cite{paulo} by Paulo Bedaque at this conference.
Here I focus on the first step in its simplest context,
that of few-nucleon systems. Bigger nuclear systems 
were discussed \cite{wolfram} by Wolfram Weise.
As we are going to see, complexity, at least in
light-nuclear structure, is largely due to fine-tuning.

It is a common situation that we are interested
in physics at a certain momentum scale $M_{lo}$, having only
partial knowledge of the underlying
theory characterized by a scale $M_{hi}\gg M_{lo}$.
The crucial feature of EFT that sets it apart from models is that 
it incorporates a consequence of the marriage between quantum 
mechanics and relativity: everything that can happen will
happen ---at least virtually. Therefore, interactions
among the low-energy degrees of freedom
take all possible forms allowed by the symmetries of the underlying
theory. In separating out some degrees of freedom as ``low-energy'',
we introduce an arbitrary ultraviolet momentum cutoff $\Lambda$.
Model independence requires that low-energy observables be independent
of $\Lambda$ ---renormalization-group (RG) invariance.
Facing an infinite
number of contributions to processes at momenta $Q\sim M_{lo}$,
the only way to make predictions is to 
set up an expansion in powers
of $Q/M_{hi}$, referred to as power counting.
Truncating at any given order,
we neglect contributions
with higher powers of $Q/M_{hi}$ and $Q/\Lambda$,
a breaking of RG invariance being unavoidable but sufficiently small
as long as $\Lambda\simge M_{hi}$. Accuracy can be improved systematically
by calculating higher-order contributions.

In the nuclear context, there are several relevant scales.
First, QCD is characterized by an intrinsic scale $M_{QCD}\sim 1$ GeV,
which sets the size of most hadron masses, for example the
nucleon mass $m_N\simeq 940$ MeV. The glaring exception is the
pion, whose mass $m_\pi\simeq 140$ MeV signals its character
as a pseudo-Goldstone boson. Pion interactions are characterized 
by the pion decay constant $f_\pi \simeq 93$ MeV $\sim M_{QCD}/4\pi$.
Moreover, low-energy pions can excite a nucleon to a delta isobar, introducing
the mass-splitting scale $m_\Delta -m_N\simeq 300$ MeV.
For simplicity I lump these three lower scales together into a 
common scale $M_{nuc}\sim 100$ MeV.
As we are going to see shortly, $M_{nuc}$ sets the scale of nuclear
interactions 
---for example, the two-nucleon effective ranges $r_{NN}\sim 1/M_{nuc}$.
Still, there is a less obvious, smaller momentum scale: 
$\aleph\sim 1/a_{NN}\sim 30$ MeV,
where $a_{NN}$ stands for some average of the two-nucleon scattering lengths.
The appearance of this scale, which we will also discuss in the next section,
determines much of the structure of light systems.

These multiple scales beg for multiple EFTs. 
Each EFT holds in a certain energy region; each, for particular values of
its parameters, {\it is} QCD in that region. 
For other values of the parameters,
these EFTs represent other possible underlying theories with 
the same symmetry pattern
as QCD, but different dynamics. 
Because they carry less and less
imprints of QCD ---the RG is a semi-group, after all--- the goal in 
building EFTs at lower and
lower energies is not to test QCD, but to understand the emergence of nuclear
structure from QCD.
The three EFTs developed
so far to deal with light nuclei are briefly discussed below.
It is of course impossible to review fifteen-plus years in twenty minutes.
I will concentrate on the thread provided by fine-tuning.
Other results and references can be found in various reviews 
\cite{reviews,reviewspiful,reviewspilesshalo,reviewshalo,reviewsatoms}.

\section{Pionful EFT}

The first nuclear EFT to be considered \cite{origins}
is also the closest to QCD: the generalization of chiral perturbation
theory ($\chi$PT) to systems with two or more nucleons.
In this EFT, $M_{lo}=M_{nuc}$ and $M_{hi}=M_{QCD}$: the low-energy degrees
of freedom are (non-relativistic) nucleons, pions, and 
(non-relativistic) delta isobars;
the symmetries, Lorentz, approximate parity, time-reversal, and chiral.
The Lagrangian is the one familiar in $\chi$PT applications to processes
with one or no nucleon, supplemented by contact interactions
among nucleons and deltas.
(The same ideas can be applied \cite{FKMvK} to other ``nuclei'',
such as possible hidden-charm ``molecules''.)

Not everything is just a trivial extension, though. There are two related, new
issues. First, contributions to nuclear amplitudes
exist where intermediate states differ
from initial states 
only in the kinetic energies of nucleons,
which are of ${\cal O}(Q^2/M_{QCD})$ ---in contrast, in hadronic processes
generic intermediate states have energies
of ${\cal O}(Q)$. 
Contributions from unitarity cuts are thus infrared enhanced 
\cite{origins} and 
typically of ${\cal O}(M_{QCD}Q/4\pi)$. 
Since one-pion exchange is of ${\cal O}(1/M_{nuc}^2)$, ladder contributions
need to be resummed in low partial waves 
---in the two-nucleon case, $l\le 2$ \cite{NTvK}.
(Inverse factors of angular momentum $l$ suppress
loop contributions in high waves.)
As a consequence, poles can appear in $T$ matrices 
for momenta $Q\sim 4\pi M_{nuc}^2/M_{QCD}\sim M_{nuc}\sim 100$ MeV,
or equivalently for binding energies 
$B\sim Q^2/m_N \sim M_{nuc}^2/M_{QCD}\sim 10$ MeV.
Nuclei thus arise naturally within EFT as a consequence of
the existence of the scale $M_{nuc}\ll M_{QCD}$ associated with pion physics.

The infrared-enhanced contributions come from states with
no pions or deltas. Hence it is convenient to define  \cite{origins} 
a potential
as the sum of Feynman diagrams that are nucleon irreducible. 
The estimate of the sizes of pion-exchange contributions to the potential
is similar to that done in $\chi$PT with at most one nucleon,
and the long-distance nuclear potential thus has \cite{origins} an expansion
in $Q/M_{QCD}$ from an increasing number of pion exchanges.
The main pieces of the long-range potential have been calculated 
\cite{origins,reviews,reviewspiful}
---for example, the two-pion-exchange three-nucleon potential
is a modification \cite{3BF}
of the popular Tucson-Melbourne potential---
and incorporated in more traditional
approaches, such as the Nijmegen phase-shift analysis
\cite{birthplace},
where they partially replace heavy-meson exchanges.

The second new issue is the size of fermion contact interactions.
Until we have a solution for low-energy QCD, the best we can do 
is dimensional analysis constrained by RG invariance:
one looks 
at the change in the contribution from an arbitrary loop 
under a natural variation in cutoff ---that is, a variation by a factor
of ${\cal O}(1)$, say 2 or 1/3--- and demands that the size of the 
contact interactions of the same form be at least of the same size. 
In $\chi$PT with at most one nucleon, where all loops are (except
in small windows of phase space)
perturbative, 
the corresponding counterterms
scale with inverse powers of $M_{QCD}$ in a reasonably simple form,
dubbed naive dimensional analysis \cite{GM}.
For two or more nucleons there is (for $l\le 2$ in two-body subsystems) 
an infinite number 
of leading-order
loops, encapsulated in a Schr\"odinger equation 
with the one-pion-exchange potential, which behaves
as $1/f_\pi^2 r^3$ at short distance $r$.
The non-perturbative renormalization of such singular potentials
can also be dealt with by looking at natural cutoff variations
in the Schr\"odinger equation.
Surprisingly, in all
waves where the singular potential is attractive, 
the counterterms
depend on $\Lambda$ in a limit-cycle-like 
fashion \cite{singular,singularcorr}, 
and their sizes are effectively driven by the scale 
appearing in the singular potential, rather than $M_{hi}$.
In the nuclear case this scale 
is $f_\pi$,
and thus certain contact interactions are larger \cite{KSW96,towards,NTvK} 
than expected on the basis of naive dimensional analysis.
Fortunately 
one can show \cite{singularcorr} that the size
of sub-leading contact interactions {\it relative to the leading ones} 
can still be estimated using naive dimensional analysis,
and, together with the sub-leading pieces of the long-range
potential, can be treated in perturbation theory in scattering amplitudes.
The end result is that 
at each order RG invariance
demands more contact interactions \cite{KSW96,towards,NTvK} than assumed 
in earlier work. 
(For a dissenting view, see Ref. \cite{germ}.)

Much attention has been dedicated to this EFT, although 
predominantly in its original version \cite{origins}:
many encouraging 
results for scattering and bound states 
have been obtained \cite{reviews,reviewspiful} and certainly many more
are to come. Here I will mention only the appearance of fine-tuning.

\begin{figure}[t]
\begin{center}
\includegraphics[height=5.0cm,width=6.5cm]{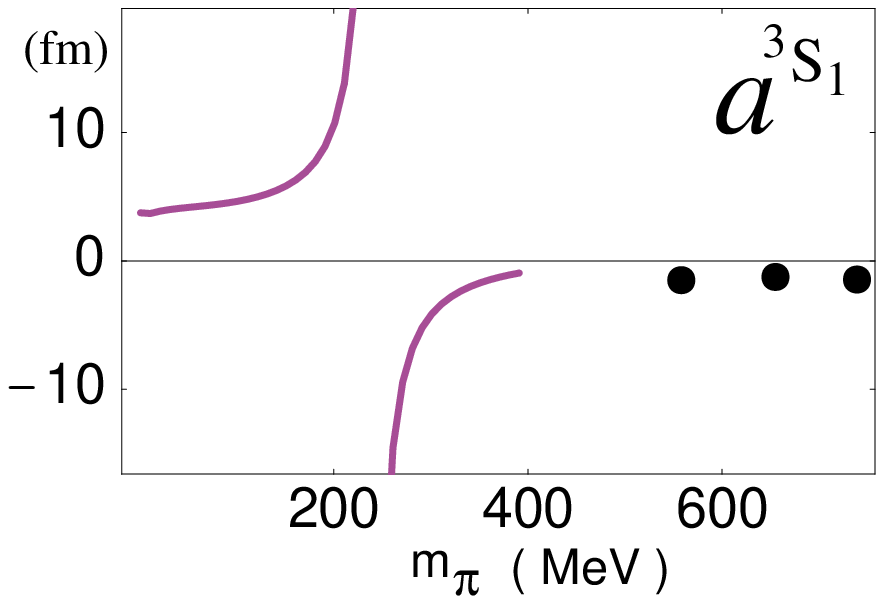}
\hspace{0.75cm}
\includegraphics[height=5.0cm,width=7.25cm]{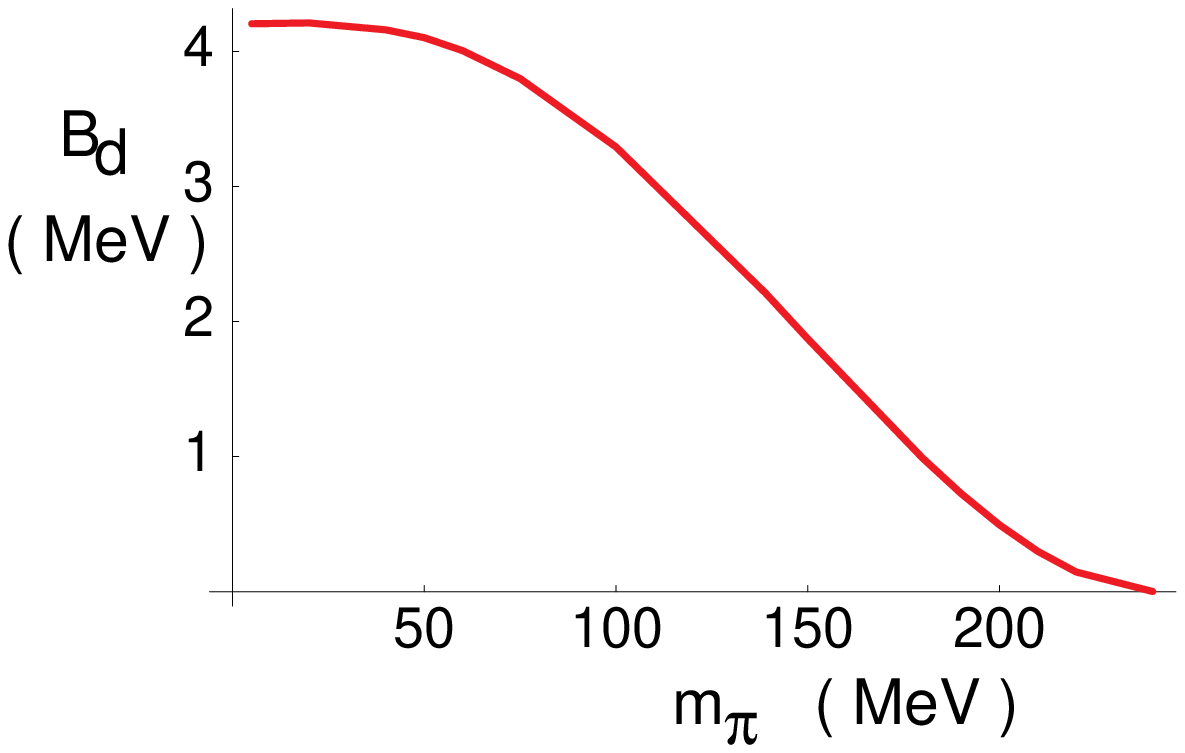}
\end{center}
\vspace*{-0pt}
\caption{Dependence
of the two-nucleon $^3$S$_1$ scattering length $a^{^3S_1}$ (left panel) 
and deuteron binding energy $B_d$ (right panel) on the pion mass $m_\pi$.
The solid lines are results  \cite{towards} from
an incomplete sub-leading-order EFT calculation,
where $m_\pi$ in the one-pion-exchange potential was varied.
The dots are quenched lattice data \cite{quenched}.
}
\label{fig1}
\end{figure}

In this EFT the pion mass can be varied, as long as $m_\pi \ll M_{QCD}$.
Thus EFT parameters can be fitted to lattice results at sufficiently
small pion masses. The idea is illustrated in Fig. \ref{fig1},
where the triplet two-nucleon scattering length 
$a_{NN}^{^3S_1}$ and the deuteron
binding energy $B_d$ in an incomplete sub-leading order \cite{towards}
are shown 
together with old, quenched lattice data \cite{quenched}.
We can see that generically, as indicated by dimensional analysis,
$a_{NN}\sim 1/M_{nuc}$ and $B_d\sim M_{nuc}^2/M_{QCD}$. 
However, a larger scattering length,
or equivalently a smaller deuteron binding energy,
arises from the seemingly accidental proximity between
the observed $m_\pi$ and the critical $m_\pi^*(M_{QCD})\sim 200$ MeV
where the deuteron goes unbound.
(Because of this fine-tuning, details of this figure
depend considerably on sub-leading
parameters that cannot at present be obtained from
experimental data ---for discussions
and newer versions of Fig. \ref{fig1}, including
unquenched lattice results, see Refs. \cite{pionmass,paulo}.)

The similarity with cold-atom systems under a varying magnetic field
is striking: QCD seems to be near a Feshbach resonance where
the role of magnetic field is played by the pion mass.
This of course does not explain why the quark masses are such that 
$m_\pi$ is close to a quantity mostly determined by $M_{QCD}$.
But it does provide a mechanism to understand the emergence of 
a new momentum scale $\aleph\sim |1-m_\pi/m_\pi^*| M_{nuc}\ll M_{nuc}$.

\section{Pionless EFT}

The emergence of bound states at momenta $Q\sim \aleph\ll M_{nuc}$ makes 
it interesting to consider an EFT where 
$M_{lo}=\aleph$ and $M_{hi}=M_{nuc}$. 
In this simpler EFT even pions and deltas can be integrated out in favor
of contact interactions among nucleons; the relevant symmetries are just 
Lorentz and approximate parity and time-reversal.
Since all that remains is contact interactions constrained by 
space-time symmetries,
the EFT can be easily generalized  
\cite{reviewsatoms} to atoms 
with large two-body scattering lengths.

Infrared enhancement is also present here, but the potential
is merely a sum of delta functions and its derivatives.
If the leading interactions scale with inverse powers of $\aleph$,
they need to be iterated and lead to bound states
with binding momenta $Q\sim \aleph$ \cite{pionless}. 
Also like in the pionful theory,
renormalization requires care because of the singular character
of the interactions. 
In the two-body system, the delta-function strengths
depend on $\Lambda$ in relatively simple ways.
Corrections
are treated in perturbation theory, and one can show \cite{pionless} that
the two-body amplitude is equivalent to the effective-range expansion.
In systems with three or more bodies, 
things are more interesting and the virtues of EFT more evident.
Some observables can be calculated to high order before
new parameters appear ---so-called low-energy theorems.
For example, nucleon-deuteron scattering in the $S_{3/2}$ channel
can be postdicted \cite{earlystooges} with QED-like precision.
In the $S_{1/2}$ channel, on the other hand,
non-perturbative renormalization requires a three-body force
already at leading order, 
which lies on a limit cycle \cite{stooges} ---just like 
two-nucleon contact interactions in the pionful theory. 
It has been conjectured 
\cite{infrared} that QCD is indeed near an infrared limit cycle.

Considerable success has been achieved
for bound states and scattering in systems with $A\le 4$ 
nucleons \cite{reviews,reviewspilesshalo},
which we can take as an indication that these systems are governed by the
fine-tuned scale $\aleph$. 
But nuclei get denser
as $A$ increases, so it is likely that at some point
binding momenta reach $M_{nuc}$.
How far in $A$ can we go with this EFT? 
Building heavier nuclei starting from inter-nucleon interactions
is in fact a major thrust of current nuclear physics.
The difficulty is that, for a given computational power,
growth in $A$ can only be achieved by a reduction in number of
one-particle states included in the calculation.
Here, like with quarks and gluons, the numerical solution requires, 
in addition to
an ultraviolet momentum cutoff $\Lambda$, also an infrared 
momentum cutoff $\lambda$
to generate a discrete one-particle basis. 
One such possibility is formulating  \cite{lattice} the EFT
in a lattice of finite size $L$ and
spacing $a$, when $\lambda\sim 1/L$
and $\Lambda\sim 1/a$.
Another, more suitable for an eventual connection
with the successful, traditional shell model,
is to formulate \cite{ionel} it instead within a harmonic oscillator of length
$b=1/\sqrt{m_N \omega}$
and a maximum number of shells $N_{max}$, when $\lambda= 1/b$ and
$\Lambda= \sqrt{(N_{max}+3/2)}/b$.
This method is known in nuclear physics 
as the No-Core Shell Model (NCSM) \cite{NCSM}.

\begin{figure}[t]
\begin{center}
\includegraphics[height=5.25cm,width=7.25cm]{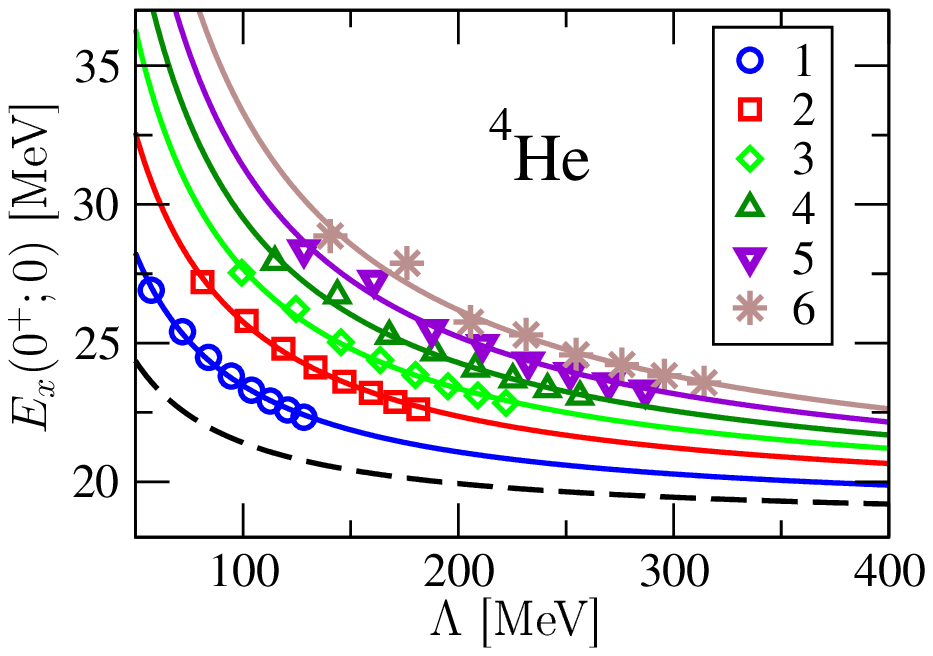}
\hspace{0.4cm}
\includegraphics[height=5.25cm,width=7.25cm]{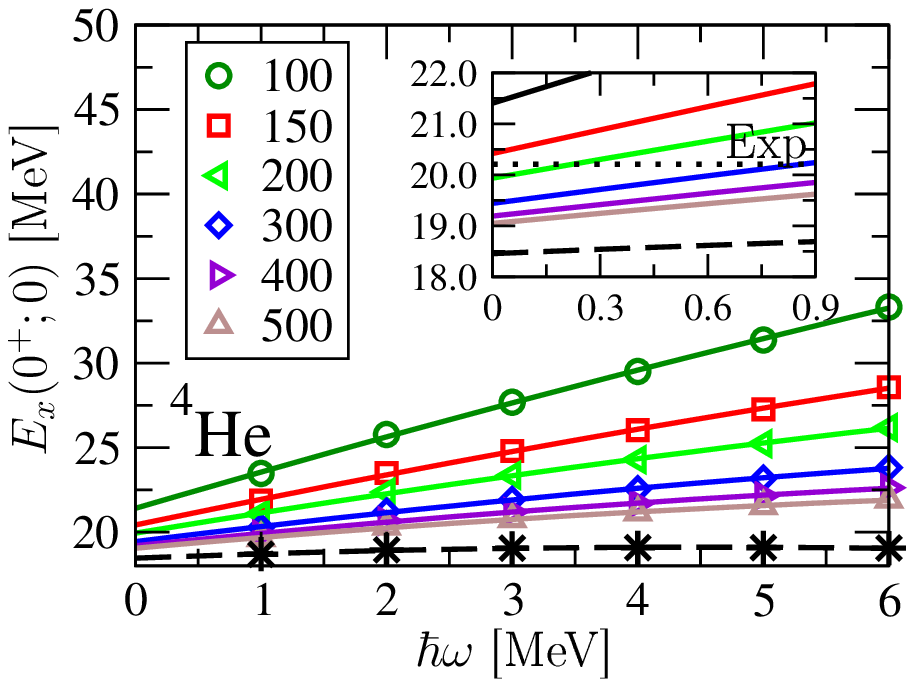}
\end{center}
\caption{
Dependence of the first 
$^4$He excitation energy $E_x(0^+;0)$ 
on the ultraviolet momentum cutoff $\Lambda$ (left panel)
and on the infrared energy cutoff $\hbar\omega$ (right panel).
The points are the leading-order results \cite{ionel}
at the values of $\hbar\omega$  (left panel)
and $\Lambda$ (right panel) given in the legends in MeV.
The (color) solid lines are simple extrapolations.
The (black) dashed lines are the limits $\hbar\omega\to 0$  (left panel)
and $\Lambda\to\infty$ (right panel).
In the insert  (right panel) we see the variation at small $\hbar\omega$,
compared to the experimental value \cite{A=4}
given by the (black) dotted line.
\label{fig2}}
\end{figure}

Using the NCSM we were able \cite{ionel} 
to push the solution of the pionless EFT beyond $A=4$.
The three leading-order parameters were fitted at 
each $\Lambda$ and $\lambda$ to the deuteron, triton, and alpha-particle
ground-state energies, and the energies of the $^4$He first-excited state 
and of the $^6$Li ground state were postdicted.
Results \cite{ionel} for the $^4$He first-excited-state energy are shown
in Fig. \ref{fig2} for various values of $\Lambda$ and $\omega$.
Convergence for $\Lambda\to\infty$
indicates correct renormalization, 
and the experimental value \cite{A=4} is reproduced within 10\% in
the limit $\omega\to 0$.
Results \cite{ionel} for $^6$Li are analogous, but they are about 30\% off
experiment.
Since this is the expected size of sub-leading terms, 
we cannot say that we are seeing a
failure of the pionless EFT. 
We are currently extending the calculation to higher orders,
will consider heavier nuclei,
and will eventually apply the same method to the pionful EFT.

\section{Halo/Cluster EFT}

Even those powerful computational methods will run out of steam
at some $A$. If we want to understand the structure of heavier nuclei
we likely need to develop other EFTs.
As a step in this direction, we can look at a class of nuclear systems,
which are sufficiently shallow for a cluster of nucleons to behave
coherently.
These are systems near thresholds for break-up into nucleons and clusters.
When nucleons orbit around a core, the
system is called a halo, but other cluster systems might have several cores.
These systems are characterized by two scales:
the energies of core excitation $E_c$ and 
of separation $E_s\ll E_c$.
Classic examples involve alpha-particle cores, for which $E_c\sim 20$ MeV:
$^6$He, where $E_s\sim 1$ MeV for break-up into two nucleons
and one alpha particle,
and the Hoyle state of $^{12}$C, where $E_s\sim 0.3$ MeV
for break-up into three alpha particles.
If $\mu$ is the reduced mass,
typically $\sqrt{\mu E_s}\sim \aleph$,
while $\sqrt{\mu E_c}\simle M_{nuc}$ is a harder scale.
Given that the alpha particle itself can, apparently, be described 
by the pionless EFT, it is not clear where this separation of scales comes 
from ---perhaps factors of $A$.

In any case, we can formulate an EFT for such halo/cluster systems,
where $M_{lo}=\sqrt{\mu E_s}$ and $M_{hi}=\sqrt{\mu E_c}$.
Its structure is analogous to the pionless EFT, with the addition 
of a field for the core. 
In order to determine the strengths of contact interactions in this EFT,
we have at first considered two-body systems:
nucleon and alpha particle \cite{Nalpha},
and two alpha particles \cite{twoalpha}.
Neither system has a bound state, but they both have non-trivial
low-energy physics in the form of resonances at $Q\sim \aleph$.
In proton-alpha and alpha-alpha scattering
the Coulomb interaction has to be incorporated explicitly.
Resonances and Coulomb can be handled in EFT,
with some technical developments.
For reviews and references, see Refs. \cite{reviewspilesshalo,reviewshalo}.

\begin{figure}[t]
\begin{center}
\includegraphics[height=6.0cm,width=8.0cm]{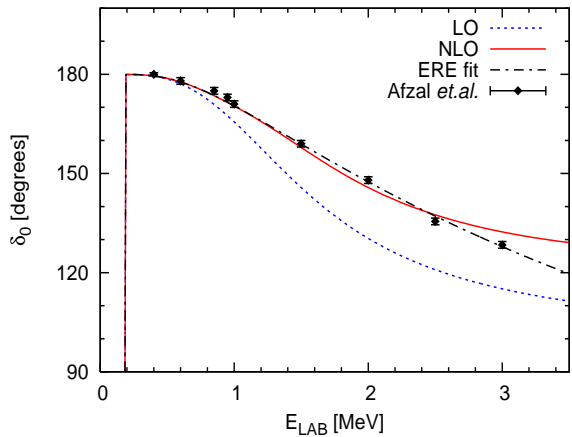}
\end{center}
\vspace*{-0pt}
\caption{Two-alpha $S$-wave, Coulomb-corrected phase shift $\delta_0$ 
as a function of the laboratory energy $E_{LAB}$. 
The EFT results  \cite{twoalpha} in leading and next-to-leading order 
are given by the (blue) dotted and (red) solid lines, respectively.
The empirical values \cite{AAA69} are the (black)
solid circles with error bars, while
an effective-range fit  \cite{twoalpha}
is given by the (black) dash-dotted line.
}
\label{fig3}
\end{figure}

In Fig. \ref{fig3} I present the leading- and next-to-leading-order 
EFT results \cite{twoalpha} for the two-alpha $S$-wave phase shift
---with purely Coulombic interactions removed---
in comparison with empirical values \cite{AAA69}.
The rapid rise at low energy represents a resonance,
which has well measured \cite{aareson} 
energy $E_R=92.07\pm 0.03$ keV and 
width $\Gamma(E_R)=5.57\pm 0.25$ eV.
In leading order the EFT has two parameters, which we fitted
to these quantities; the energy dependence of the
scattering amplitude is a postdiction.
In next order, there is a third parameter which we fitted 
to improve the energy dependence.
As one can observe, a good, converging description
of this system is achieved. 

What is remarkable is the amount of fine-tuning required for
such a description. The Coulomb scale for both scattering length
and effective range is half the Bohr radius, about 2 fm.
Coulomb indeed provides an energy dependence of the expected size;
however, the strong interaction produces an energy dependence
that cancels it within 10\% \cite{twoalpha}.
In order to produce a resonance 
at the low energy where it is observed, another factor, 100, appears and
the energy-independent
part of the amplitude has to have a scattering length
$a_{\alpha\alpha}=-(1.92\pm 0.09)\cdot 10^3$ fm \cite{twoalpha}.
We are thus effectively looking at a fine-tuning by a factor of 1000!
Note that it is different from the one seen previously,
which did not involve Coulomb.
The Hoyle state of $^{12}$C has long been considered 
\cite{3alpha} an example
of fine-tuning, but the fine-tuning is already mind-boggling in
the simpler two-alpha system. It remains to be seen if in the EFT  framework
the Hoyle state arises naturally from the two-alpha interactions 
found in the two-alpha system.

\section{Conclusion}

Many of us are prejudiced to think that fine-tuning does not happen.
However, I have argued that 
at least part of the complexity found in nuclear physics is rooted 
in fine-tuning. 
In the process of uncovering this fine-tuning, quite a number
of other cool features surfaced in nuclear EFTs, such
as limit cycles and wide universality.

As discussed above, there is a natural reason for nuclei to be shallow 
with respect to the intrinsic QCD scale ---natural in the sense that 
it appears in the pionful EFT 
from the scales of spontaneous chiral symmetry breaking.
Yet,
the deuteron and other light nuclei are somewhat closer to being unbound
than expected, which can be traced to an apparent accident:
the value of the pion mass is close to a critical value.
One can go a long way in explaining properties of
light nuclei by incorporating this fine-tuning in the pionless EFT.
In this context, nucleons are just cold fermions near a Feshbach resonance.

As if that was not enough, in alpha-alpha scattering in the halo/cluster EFT 
further fine-tuning
seems to arise 
between strong and electromagnetic interactions:
not one in three or four, as in the pion mass, but one in a thousand!
Again the fine-tuning can be incorporated in the EFT,
but its origin remains elusive.

EFT provides a context for emergence. And in QCD quite a lot emerges...

\vspace{.5cm}
\noindent
{\bf Acknowledgments.} 
I thank my collaborators, particularly Renato Higa and Ionel Stetcu,
for their insights into much of the work reported here.

\end{document}